\documentclass[english,twocolumn,showpacs,prl,amsmath,amssymb,superscriptaddress,floatfix]{revtex4}
\usepackage{graphicx}
\usepackage{color}

\newcommand{\tr}{{\rm Tr\thinspace}}

\newcommand{\expect}[1]{\ensuremath{\left\langle{#1}\right\rangle}}

\begin{document}
\title{Converting a real quantum spin bath to an effective classical noise acting on a central spin}
\author{Wayne M. Witzel}
	\affiliation{Sandia National Laboratories, Albuquerque, New Mexico 87185 USA}
\author{Kevin Young}
	\affiliation{Sandia National Laboratories, Livermore, California 94550 USA}
\author{Sankar Das Sarma}
	\affiliation{University of Maryland, College Park, Maryland 20742-4111, USA}

\begin{abstract}\noindent
We present a cluster expansion method for approximating quantum spin-bath dynamics in terms of a classical Gaussian stochastic process.  The cluster expansion produces the two-point correlation function of the approximate classical bath, permitting rapid evaluation of noise-mitigating quantum control strategies without resorting to computationally intensive dynamical decoupling models.  Our approximation is valid for the wide class of models possessing negligible back-action and nearly-Gaussian noise.  We study several instances of the central spin decoherence problem in which the central spin and randomly-located bath spins are alike and dipolarly coupled. For various pulse sequences, we compare the coherence echo decay computed explicitly quantum mechanically versus those computed using our approximate classical model, and obtain agreement in most, but not all, cases.
We demonstrate the utility of these classical noise models by efficiently searching for the 4-pulse sequences that maximally mitigate decoherence in each of these cases, a computationally expensive task in the explicit quantum model.
\end{abstract}

\maketitle\noindent
Spin echo decay, which operationally defines the dephasing/decoherence time $T_2$, is an important measure characterizing the viability of prospective qubit realizations.  For solid-state spin qubits, this echo decay may be attributed to the dynamical flip-flopping of impurity spins surrounding the central qubit spin.  Cluster expansion techniques \cite{witzel_quantum_2005, witzel_quantum_2006, yao_theory_2006, yang_quantum_2008, yang_quantum_2009, cywinski_electron_2009, cywinski_pure_2009, witzel_quantum_2012}, which simulate the microscopic quantum dynamics of the spin-bath system, have proven exceptionally reliable in quantitatively reproducing and predicting measured spin echo decays ~\cite{witzel_decoherence_2007, witzel_electron_2010, cywinski_electron_2009, cywinski_pure_2009, bluhm_dephasing_2010, george_electron_2010, balian_measuring_2012} and in the evaluation of dynamical decoupling strategies~\cite{yao_restoring_2007, witzel_multiple-pulse_2007, witzel_concatenated_2007, lee_universal_2008, zhao_anomalous_2011, zhao_decoherence_2012}, which extend the coherence time of the central spin through the application of precisely timed pulse sequences.  These studies, however, are computationally intensive and must be repeated for each pulse sequence under consideration.  

In contrast, the echo decay of a quantum spin subjected to \emph{classical} Gaussian noise may be computed extremely efficiently in terms of filter functions and the noise correlation function~\cite{cywinski_how_2008,ascheron_electron_2009,young_qubits_2012}.  This efficiency facilitates, for example, optimal control calculations that would be computationally intractable on a fully quantum model. Such considerations lead us to question under what circumstances a quantum spin bath Hamiltonian, consisting of many interacting impurity spins, may be well approximated as a classical stochastic noise.  

A semiclassical stochastic noise model should well approximate the dynamics of a fully quantum model if two conditions are met: (i) the bath dynamics are independent of the central spin state, \emph{i.e.} back action effects are insignificant, so the bath effects on the central spin appear classical and (ii) the effective noise is approximately Gaussian, so is characterized completely in terms of its two-point correlation function, and is therefore amenable to filter function techniques.  In this paper, we shall consider a central spin decoherence problem in which the central spin and randomly-located bath spins are alike and dipolarly coupled, such as an electron spin in an electron spin bath.   
The large, identical spin bath possessed by this model is very likely to satisfy both of the above conditions: the first because the bath and central spins are alike, so the state of the central spin is unlikely to drastically affect the bath dynamics; the second because the spin bath resembles as a large collection of two-level fluctuators which combine to yield Gaussian statistics, as implied by the central limit theorem.   This model has been extensively studied with cluster expansion methods in previous work~\cite{witzel_quantum_2012}, and evidence ~\cite{hanson_coherent_2008, dobrovitski_decay_2009, de_lange_universal_2010, witzel_quantum_2012, wang_spin_2013} suggests that the bath effects are well represented as a classical Ornstein-Uhlenbeck (O-U) stochastic process.  

As the only stochastic process which is Markovian, Gaussian and stationary, O-U noise represents an idealized approximation to the quantum dynamics.  In this paper we extend this semiclassical approximation, generalizing the classical stochastic process so that its correlation function matches that of the fully quantum model, which we compute using a modified cluster expansion.  We then compare the resulting spin echo decays against those computed directly with a fully quantum mechanical treatment.  Remarkable agreement is shown for most instances of the randomly distributed bath spins.  After demonstrating the broad validity of the semiclassical models, we further illustrate their utility through the construction of optimally noise-mitigating pulse sequences.  

The free evolution Hamiltonian of our problem is:
\begin{equation}
\label{Eq:Ham}
\hat{\cal H} = \sum_i \mu_B g_i B_i \hat{S}^{z}_i + \mu_B^2 \sum_{j>i} g_i g_j
\hat{\bf S}_i \cdot {\bf D}({\bf R}_i -  {\bf R}_j) \cdot \hat{\bf S}_j,
\end{equation}
written in atomic units ($\hbar = 1$ and $1 / 4 \pi \epsilon_0 = 1$)
where $\hat{\bf S}_i$ are spin operators for the spin-1/2 particles,
$\mu_B$ is the Bohr magneton, $g_i$ is the $g$-factor of the
$i$th electron, $B_i$ is the externally applied magnetic field at each electron site, and ${\bf D}({\bf r})$ is a tensor to characterize dipolar
interactions and is defined by
\begin{equation}
\label{Eq:Dip}
D_{\alpha, \beta}({\bf r}) =
\left[ \frac{\delta_{\alpha \beta} - 3 r_{\alpha} r_{\beta} / {\bf r}^2}{{\bf r}^{3}}\right],
\end{equation}
with $\alpha, \beta = x, y, z$.
$\delta_{\alpha \beta}$ is the Kronecker delta and $r_{\alpha}$ is the $\alpha$
vector component of ${\bf r}$.  Our convention is to index the central spin as $i=0$.  We shall investigate a number of randomly generated spatial configurations of electron spins at average concentration $10^{13}$ cm$^{-3}$ and with
$g_i=2$.  We assume the limit in which $B_i$ is large and equal amongst the bath spins but not necessarily the central spin
\footnote{Ref.~\cite{witzel_quantum_2012} also considered the scenario in which the central spin is resonant with the bath spins; in that case, the short time echo decay is dominated by direct flip-flops with individual bath spins (1-cluster contributions).},
permitting a secular approximation of the interbath dynamics. That is, processes must conserve the net polarization of the bath spins; bath spins may flip-flop with each other but not the central spin.  Such a situation may arise for an addressable qubit tuned off resonance with the bath spins.  Under the secular approximation in the rotating frame, the effective Hamiltonian is
\begin{equation}
\label{Eq:HeffOffRes}
\hat{\cal H}_{\mbox{\scriptsize eff}} = \sum_{i, j > 0} b_{i, j} \hat{S}_i^+ \hat{S}_j^-
- 2 \sum_{i, j} b_{i, j} \hat{S}_i^z \hat{S}_j^z.
\end{equation}
where $b_{i, j} = -g_i g_j \mu_B^2 \hbar (1 - 3 \cos^2\theta_{ij})/(4 R^3_{ij})$,  $\theta_{ij}$ is the polar angle of the displacement vector connecting spin $i$ and spin $j$ with respect to the $\hat z$ unit vector (the direction of applied  magnetic field), and $R_{ij}$ is its length.  

We are interested in computing the spin echo decay, which is proportional to the expectation value of the coherence operator on the central spin as a function of time, $\langle \sigma_{+}(t) \rangle = \tr(\sigma^+_0 U(t) \rho(0) U(t)^\dagger)$, where the initial state of the bath is taken to be thermal at some relevant temperature. In what follows, we present a method for approximating this expectation value by constructing an effective semiclassical Hamiltonian for the central spin 
$\hat{\cal H}_{\rm cl}(t) = \mu_B g_0 B(t) \hat{S}^{z}_0$,
where the effective magnetic field $B(t)$ is a classical stochastic variable whose action approximates that of the quantum operator $\hat{B}_z = - 2 \sum_{j>0} b_{0, j} \hat{S}_j^z$.  We permit the system to be subject to $\pi$-pulse control about the $x$ axis.  In a toggling frame \cite{young_qubits_2012,ascheron_electron_2009}, each $\pi$ pulse causes the effective field felt by the central spin to flip, leading to a Hamiltonian $\tilde{H}(t) = \mu_B g_0 y(t) B(t) \hat{S}^{z}_0,$ where $y(t) = \pm 1$ changes its sign whenever a $\pi$ pulse is applied.  If the fluctuating field is Gaussian \cite{jacobs2010stochastic}, then the coherence decay may be calculated directly as
\begin{eqnarray}
	\label{eq:sigplus}
	\langle \sigma_{+}(t) \rangle 
		&= \frac{1}{2} \exp{\left(-\frac{\mu_B^2 g_0^2}{4}\int_0^{t} du~C(u) F_t(u) \right)}, \\
	\label{eq:filterfn}
	F_t(u) 
		&= \int_u^{2t - u} dv\;y\left(\frac{v+u}{2}\right) y\left(\frac{v-u}{2}\right).
\end{eqnarray}
where $F_t(u)$ is a time-domain filter function\cite{young_qubits_2012} describing the action of the control pulses and $C(t)=\expect{B(t)B(0)}$ is the correlation function of the effective classical field.  We choose this correlation function to be equal to that of the fully quantum model, $C_{\rm Q}(t)=\expect{\hat{B}_z(t)\hat{B}_z(0)}$, with $\hat{B}_z(t)$ the operator in the Heisenberg picture.  The filter function is efficiently computable, so knowledge of the correlation function is sufficient to rapidly determine the coherence remaining in the system after any sequence of $\pi$ pulses.  We  compute the quantum correlation function by using a variant\cite{witzel_quantum_2012} of the cluster correlation expansion \cite{yang_quantum_2008, yang_quantum_2009} (CCE).

The original CCE assumed the coherence decay (or any observable quantity), $L = \langle \sigma_{+}(t) \rangle$, could be decomposed as a product of contributions, $L = \prod_{\cal S} \tilde{L}_{\cal S}$, from each subset, $\mathcal{S}$, of bath spins.  The modified contributions, $\tilde{L}_{\cal S}$, are then defined implicitly through the relation, $\tilde{L}_{\cal S} = L_{\cal S} / \prod_{{\cal C} \subset {\cal S}} \tilde{L}_{\cal C}$, where the product is taken over all subsets, $\cal C$ of the set $S$ of bath spins  (we shall refer to subsets of $n$ bath spins as \emph{$n$-clusters}).  Each of the unmodified contributions ${L}_{\cal S}$ may be computed by exactly solving the dynamics of a system of bath spins $\mathcal{S}$ much smaller than the original problem.  By decomposing the observable in this manner, the solution may be successively approximated by including relevant clusters of increasingly large size.  A Dyson series expansion \cite{Sakurai1993} implies that only small clusters should be relevant to the short-time dynamics, with clusters of increasing size becoming more important with increasing time.  This small-cluster approximation has been quite successful, showing remarkable agreement with experimental data in a number of  previous studies~\cite{witzel_decoherence_2007, witzel_electron_2010, bluhm_dephasing_2010, george_electron_2010, balian_measuring_2012}.  Reference~\cite{witzel_quantum_2012} refined the performance of the CCE by providing heuristics to select and evaluate a subset of the clusters that are more likely to contribute (e.g., ones that are strongly interacting with each other), a strategy we employ in this work.  We note that the CCE is related to a linked clusters perturbation expansion~\cite{saikin_single-electron_2007} but is more convenient to evaluate in an automated way.  

Unfortunately, for a sparse bath of like spins, the CCE suffers numerical instability issues in the evaluation of $\tilde{L}_{\cal S}$ due to the occasional division by small numbers.  The physical system we consider here can be particularly vulnerable to this problem because the decoherence rate is strongly dependent on the initial state of the bath.  Consider, for example, a completely polarized bath in which there are no flip-flopping spins and therefore no nontrivial decoherence.  Other states exhibit the opposite extreme.  Because of this diversity, for times at which the expected coherence has not yet decayed significantly, there may exist spin configurations for which the expected coherence is zero, implying the $\tilde{L}_{\cal S}$ formula will involve a division by zero.  In Ref.~\cite{witzel_quantum_2012} we presented as a solution to this problem a highly technical variation of the CCE that we called interlaced spin averaging, in which each  evaluation of $\tilde{L}_{\cal S}$ was averaged over bath spin states in a relatively efficient manner (which is the fairly technical aspect), thus removing the numerical instability.  Here we present a far simpler solution in which we 
formulate $L$ as a \emph{sum} of contributions rather than a product.
Thus,
\begin{equation}
\label{Eq:additiveExpansion}
L_{\cal S} = \sum_{{\cal C} \subseteq {\cal S}} \tilde{L}_{\cal S},~ \tilde{L}_{\cal S} = L_{\cal S} - \sum_{{\cal C} \subset {\cal S}} \tilde{L}_{\cal C}.
\end{equation}
We also redefine $L$ as $\langle \sigma_{+}(t) \rangle - 1$ so it is a proper expansion about $t=0$~\footnote{Since $L(t=0)$ defined in this way is identically zero, contributions from each cluster are relative to this zero point and the expansion converges most rapidly near $t=0$; thus it is an expansion about $t=0$.}.
As in the original CCE, this is exact in the limit that all cluster contributions are included (but without any division by zero pathology).  We find the convergence behavior is different but comparable to the multiplicative version with interlaced spin averaging.

\begin{figure}
\includegraphics[width=\linewidth]{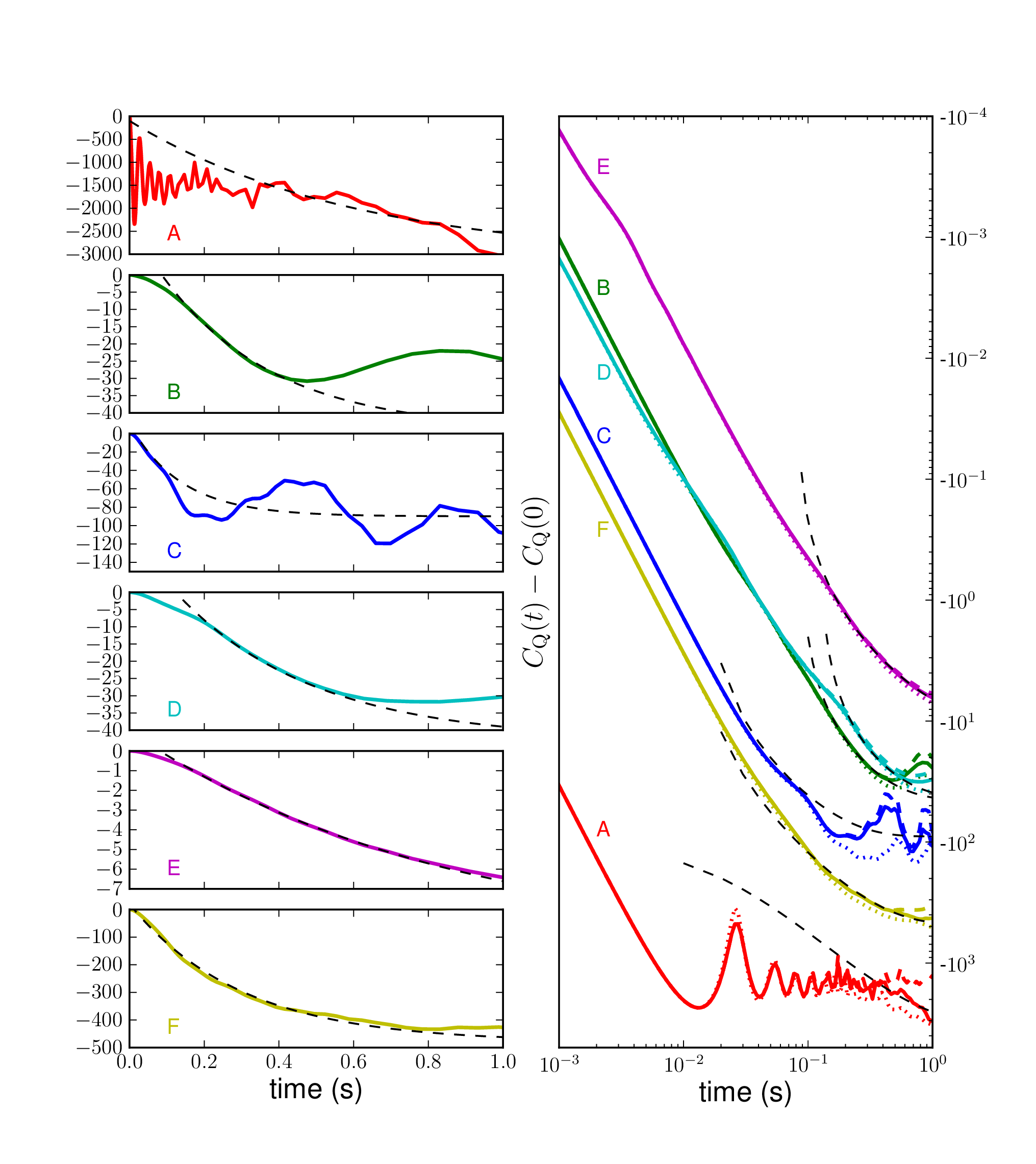}
\caption{\label{Fig:CorrFns} (Color online) Relative correlation function calculation results, $C_{\rm Q}(t)-C_{\rm Q}(0)$ in $(\mbox{rad}/\mbox{s})^2$, for the five cases of bath spatial configurations labeled A-F studied in Ref.~\cite{witzel_quantum_2012} with linear scales (left) and logarithmic scales (right).  The $C_{\rm Q}(0)$ values are $55594$, $14.3$, $59.8$, $19.1$, $5.93$, and $287$ $(\mbox{rad}/\mbox{s})^2$, respectively.
Black dashed curves are O-U type of the form $A \exp{(-B t)} + C$, loosely fitting the calculations over some respective time ranges.  Thick (and colored) solid, dashed, and dotted curves on the right plot are 2-cluster, 3-cluster, and 4-cluster results, respectively.}
\end{figure}

To directly compute correlation functions, we employ this additive form of the cluster expansion Eq.~\eqref{Eq:additiveExpansion}.  The only change is taking the quantity of interest to be the relative~\footnote{This is again defined as an expansion about $t=0$.  Note that $C_{\rm Q}(0)$ is easy to compute separately and add back in but irrelevant for any refocusing pulse sequences, such as Hahn echoes.} correlation function $L = C_{\rm Q}(t) - C_{\rm Q}(0)$.  
We estimate the average over initial spin states of the bath by taking random samples of up/down product states and compute the cluster expansion each initial spin state separately.
In Fig.~\ref{Fig:CorrFns}, we show results of these relative correlation function calculations for five cases of different random spatial locations of bath spins, the same cases labeled A-F in Ref.~\cite{witzel_quantum_2012}.  The right plot displays all cases together with 2-cluster, 3-cluster, and 4-cluster results.  Each case demonstrates convergence in time with respect to cluster size.  
Short time dynamics are dominated by 2-cluster contributions, with higher-order contributions becoming necessary with increasing time.
Case C illustrates this most clearly.

This figure also makes comparisons with exponential-like decay of O-U correlation functions.   In Fig.~6 of Ref.~\cite{witzel_quantum_2012}, we fit Hahn echo decay results of these cases with the form $\exp{(-t^3)}$ as a confirmation of the O-U noise approximation.  These were good fits on the time scale of the initial substantial echo decay (out to the first $25\%$ to $50\%$ of the decay).  With exception to case A, which is an unusual case as we shall see, the O-U noise approximation fits our correlation function results well on the same corresponding time scales.  For short times, however, we see in the right plot that the correlation functions are of the form $\alpha-\beta t^2$, as predicted by perturbation theory.

\begin{figure}
\includegraphics[width=\linewidth]{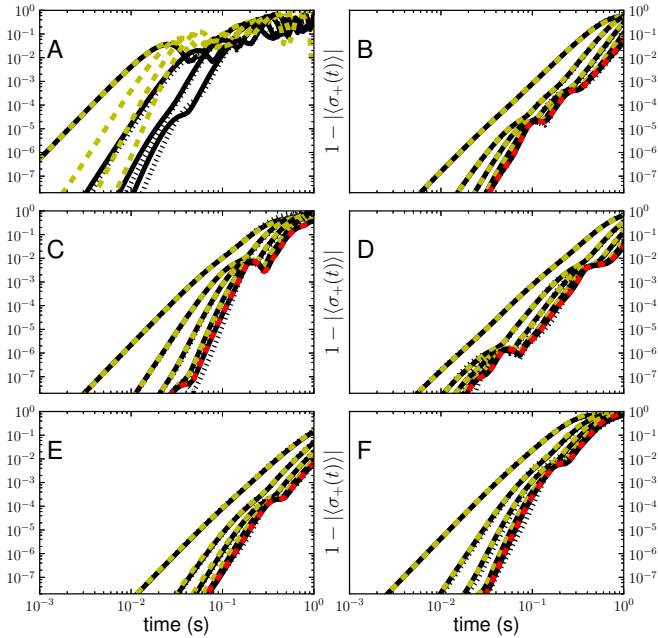}
\caption{\label{Fig:EchoComparisons} (Color online) Comparison of echo decays (plotted as $1 - $~Echo versus total pulse sequence time on a log-log scale) calculated directly using the CCE versus via the correlation functions from Fig.~\ref{Fig:CorrFns} for cases A-F.  The solid black curves are direct calculation results.  The yellow dashed curves are derived from correlation functions.  Both include up to 4-cluster contributions.  Each case shows results for 1- to 4- pulse UDD sequences (UDD1 is a Hahn echo).  Initial coherence behavior improves with an increased number of pulses; that is, the 1-4 pulse curves are seen left to right.  Additionally for cases B-F we show echo decay for optimized 4-pulse sequences that perform slightly better than UDD4; the red dashed curves are the correlation function derived results for these pulse sequences, in excellent agreement with direct calculations in solid black.   Dotted black curves show direct 2-cluster results; we find that corrections from larger clusters are important, even at short times, for some of the multipulse sequences.}
\end{figure}

While this correlation function is well defined as a quantum mechanical expectation value and we believe the cluster expansion is working well to successively approximate this quantity, whether or not the approximating classical model is sufficient for calculating echo decays is a separate question.  As discussed earlier, our semiclassical approximation relies on two assumptions: (i) minimal back-action and (ii) approximately Gaussian noise.  To verify that these assumptions hold, we compare the echo decays computed from the correlation function to echo decays computed directly using the CCE in Fig.~\ref{Fig:EchoComparisons}.  We do this on cases A-F using Uhrig dynamical decoupling (UDD)\cite{uhrig_concatenated_2009} sequences with 1-4 pulses (1-pulse UDD is a Hahn echo).

For case A, we only see agreement for the Hahn echo sequence in Fig.~\ref{Fig:EchoComparisons}.  For all other cases, we see excellent agreement for all pulse sequences.  The failure of the semiclassical model for case A implies a breakdown of one of the above assumptions.
In the absence of back-action, our echo decay results should scale in a simple manner as we reduce the central-spin gyromagnetic ratio.  That is what we find, implying that it is the Gaussian noise assumption that is violated in case A.  Indeed, by looking at the distribution of correlation function contributions for different initial spin states at various times, we see that case A exhibits multiple peaks.  The other cases typically exhibit single peaks well approximated as Gaussian.  Figure 1 of Ref.~\cite{witzel_quantum_2012} reveals that case A, by happenstance, has one particular bath spin that has a conspicuously strong interaction with the central spin and we find that the dynamics of this spin dominates the noise, which takes on non-Gaussian random telegraph character \cite{jacobs2010stochastic}.  This spin interacts strongly with two other nearby bath spins and this small system dominates the initial decoherence of the central spin.  The central limit theorem does not apply and the noise cannot be approximated as Gaussian, thus making case A a rather nongeneric special situation.  As a quick test, the Gaussianity assumption may be verified by computing \emph{four}-point correlation functions and computing the deviation from the Wick's theorem prediction.

\begin{figure}
\includegraphics[width=\linewidth]{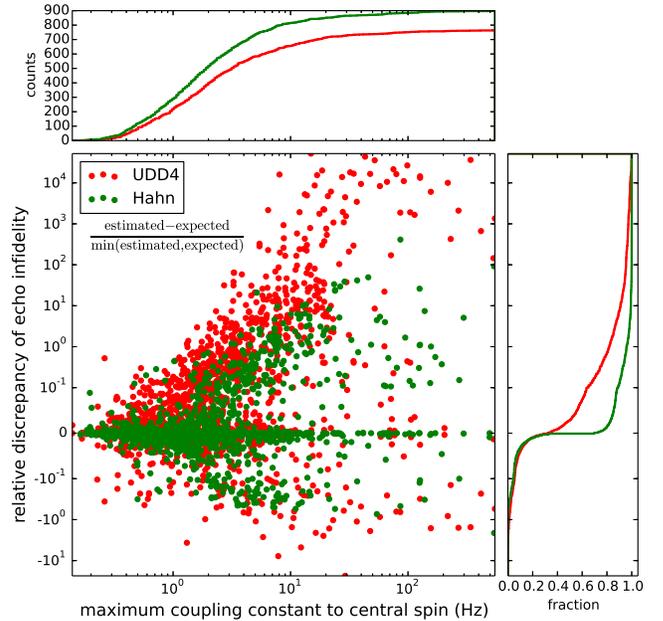}
\caption{\label{Fig:scatterComparison} Scatter plot of the extreme values for the relative discrepancy of echo infidelity (1 - echo) estimates versus the maximum coupling constant to the central spin.  Hahn echo (UDD1) and UDD4 cases are represented from about 900 and 750 random spatial configurations respectively.  The relative discrepancy is computed as (estimated - expected)/min(estimated, expected) where ``estimated'' uses the correlation function and ``expected'' uses the direct CCE approach.  Positive values correspond with pessimistic estimates.  Relative discrepancies shown are the two extremes (per spatial configuration) over $100$ logarithmically spaced time points between $1$~ms and $0.3$~s, skipping infidelities below $10^{-7}$ where the numerical precision is suspect.  These are plotted on a symmetric positive and negative logarithmic scale with a linear region between $\pm 0.1$.  The top figure shows cumulative counts of instances versus maximum coupling constants.  The right figure shows cumulative probability distributions of relative discrepancies.}
\end{figure}

We performed a statistical study of about 900 and 750 random instances respectively for Hahn echo and UDD4 which shows a general correlation between atypical instances (like case A) and the presence of particularly strongly coupled (nearby) bath spin.  Figure~\ref{Fig:scatterComparison} shows scatter plots of relative discrepancies of echo infidelity versus maximum coupling to the central spin.  Relative discrepancies are comparisons between correlation function results and direct CCE evaluations of the echo infidelities (1- echo).  Instances where the maximum coupling between central and bath spins is small will likely have noise dominated by many similar bath spins and will tend to be Gaussian.  The non-Gaussian cases are ones in which the noise is dominated by a small number of spins and thus more likely to have a large maximum coupling.  Indeed, we find that discrepancies increase in magnitude as the maximum coupling increases, confirming that this is a good signature of the degree to which the noise can be expected to be Gaussian.  The Hahn echo is much better behaved indicating that the atypical behavior tends to manifest itself with increasing dynamical decoupling as we saw in case A above (see Fig.~{\ref{Fig:EchoComparisons}}).  Also note that the correlation function estimates tend, strongly, to error on the side of being pessimistic (positive relative discrepancies).  These results over a large number of random instances were computed using an automated convergence strategy for selecting clusters with a convergence criteria of 10\% variation in the logarithm of the contribution, per cluster size, to the echo infidelity.  Because some instances failed to converge at the end of the allotted 48 hour run time on 128 processors, there may be some bias with these ``harder'' instances being underrepresented.

\begin{figure}
\includegraphics[width=\linewidth]{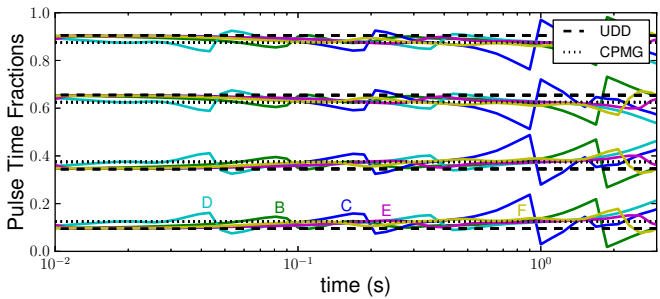}
\caption{\label{Fig:Opt4Pulse} (Color online) Optimal, symmetric, 4-pulse refocusing sequences for cases B-F found efficiently using the correlation functions of Fig.~\ref{Fig:CorrFns}.  The curves indicate the fractional times of $\pi$ pulses as a function of the total pulse sequence time.  For comparision, we show the $4$-pulse UDD and Carr-Purcell-Meiboom-Gill sequence as dashed and dotted lines respectively.}
\end{figure}

In most cases the correlation function is an adequate and convenient characterization of the noise.  Where it is valid to use, we can use it to compute the echo decay for any pulse sequence very efficiently.  
We offer a simple demonstration of this by optimizing over the symmetric, 4-pulse, refocusing sequences (parameterized by a single number).  We show these optimal sequences as a function of total pulse sequence time for cases B-F in Fig.~\ref{Fig:Opt4Pulse}.  We leave out the nongeneric case A since the correlation function is not a sufficient characterization of its noise.  In Fig.~\ref{Fig:EchoComparisons} we see that these optimal sequences perform slightly better than UDD4.  While the improvement is not substantial, it demonstrates the utility of a correlation function description of the noise.

In conclusion, we have presented a method for deriving a semiclassical correlation function directly from a microscopic quantum spin-bath model using a cluster expansion approach.  We have applied this approach to the problem of a dipolarly-interacting system of sparse, like spins finding that the correlation function is an adequate characterization of the noise for most, but not all, cases of this problem.  Some nongeneric instances are dominated by small system dynamics producing non-Gaussian noise.  We estimated the statistical likelihood for an instance to be atypical and revealed a correlation between non-Gaussian noise behavior and the presence of a bath spin that has a particularly strong interaction with the central spin.   We establish that correlation functions, where applicable, can be used to efficiently find optimal pulse sequences.  With this tool, we can learn about the difference in the performance of generic pulse sequences versus optimized pulse sequences that are tailored to specific qubit environments.  Our explicit conversion of a quantum bath Hamiltonian to an effective classical noise description allows for a tremendous improvement in the efficiency of evaluating various control pulse sequences to preserve system coherence, enabling optimized quantum error correction protocols for spin qubit architectures.

We thank and acknowledge Rogerio de Sousa, Robin Blume-Kohout, Toby Jacobson, Erik Nielsen, Rick Muller, Malcolm Carroll, and particularly {\L}ukasz Cywi{\'n}ski for valuable discussions and contributions.  We further acknowledge the IARPA QCS program whose support for WMW and KY initiated our line of inquiry.  
Sandia National Laboratories is a multi-program laboratory managed and operated by Sandia Corporation, a wholly owned
subsidiary of Lockheed Martin Corporation, for the U.S. Department of Energy’s National Nuclear Security Administration
under contract DE-AC04-94AL85000. The work at the University of Maryland is supported by LPS-CMTC and IARPA.

\bibliography{Refs}
\end{document}